\documentclass[12pt]{iopart}

\expandafter\let\csname equation*\endcsname\relax
\expandafter\let\csname endequation*\endcsname\relax
\usepackage{amsmath}
\usepackage{cite}
\usepackage{graphicx}
\usepackage{graphicx}

\usepackage{multirow}
\usepackage{upgreek}
\usepackage{textcomp} 
\usepackage{gensymb}
\usepackage{xcolor} 
\usepackage{csquotes} 
\usepackage{hyperref} 
\usepackage{caption, tabularx}
\usepackage{enumitem, ragged2e}
\makeatletter
\newcommand*{\compress}{\@minipagetrue}
\makeatother
\usepackage{siunitx} 
\usepackage{indentfirst}
\newcommand{\cc}[1]{\textcolor{black}{#1}}

\begin{document}

\title[Machine learning for FLIM]{Machine learning for faster and smarter fluorescence lifetime imaging microscopy}

\author{Varun Mannam$^1$, Yide Zhang$^{1,2}$, Xiaotong Yuan$^1$, Cara Ravasio$^1$ and Scott S. Howard$^1$*}
\address{$^1$Department of Electrical Engineering, University of Notre Dame, Notre Dame, IN 46556, USA}
\address{$^2$Caltech Optical Imaging Laboratory, Andrew and Peggy Cherng Department of Medical Engineering,  California Institute of Technology, Pasadena, CA 91125, USA}
\ead{showard@nd.edu}
\vspace{10pt}
\begin{indented}
\item[]March 2020
\end{indented}

\begin{abstract}
Fluorescence lifetime imaging microscopy (FLIM) is a powerful technique in biomedical research that uses the fluorophore decay rate to provide additional contrast in fluorescence microscopy. However, at present, the calculation, analysis, and interpretation of FLIM is a complex, slow, and computationally expensive process. Machine learning (ML) techniques are well suited to extract and interpret measurements from multi-dimensional FLIM data sets with substantial improvement in speed over conventional methods. In this topical review, we first discuss the basics of FILM and ML. Second, we provide a summary of lifetime extraction strategies using ML and its applications in classifying and segmenting FILM images with higher accuracy compared to conventional methods. Finally, we discuss two potential directions to improve FLIM with ML with proof of concept demonstrations.
\end{abstract}
\vspace{2pc}

\section{Introduction}
\subsection{Motivation} \label{moto}
Fluorescence microscopy is an essential tool in biological research that enables both structural and functional molecular imaging of biological specimens \cite{FluorescenceMicroscopy}. Fluorescence lifetime imaging microscopy (FLIM) augments fluorescence microscopy by measuring the effects of biological micro-environment on fluorescence decay rate \cite{FLIM2007}. While powerful, extraction, and interpretation of the lifetime information is a complicated and time-consuming process, as the raw FLIM data are multi-dimensional, and the FLIM algorithms are computationally intensive.  

Recently, machine learning (ML), particularly deep learning \cite{goodfellow2016deep}, has gained significant traction for its excellent performance in processing complex multi-dimensional data. Whereas multiple ML methods have been successfully utilized to extract fluorescence lifetime information, a comprehensive study of the ML techniques applied to FLIM is still missing in the literature. In this topical review, we aim to fill this gap by reviewing the articles reporting ML methods for FLIM applications. We also envisage two potential applications of utilizing ML techniques to extract lifetime information with a small FLIM dataset.

\subsection{Fluorescence microscopy} \label{fm}
Fluorescence microscopy is one of the most powerful and versatile optical imaging techniques in modern biology as it enables the study of cellular properties and dynamics \textit{in vitro} and \textit{in~vivo}. Fluorescence microscopy illuminates a specimen to excite fluorophores within cells and tissue and then collects the emitted fluorescence light to generate images. To enhance contrast and identify specific features (like mitochondria), fluorescence labeling can be introduced. Fluorescent labeling methods include genetic labeling using fluorescent proteins \cite{proteins}, and chemical labeling by targeting species with antibodies \cite{FluorescentLabeling}. By adequately filtering the different wavelengths of the emitted light, the labeled features can be imaged and identified \cite{FluorescenceMicroscopy}.

\begin{figure}[!t]
	\centering
	\includegraphics[width=0.8\linewidth]{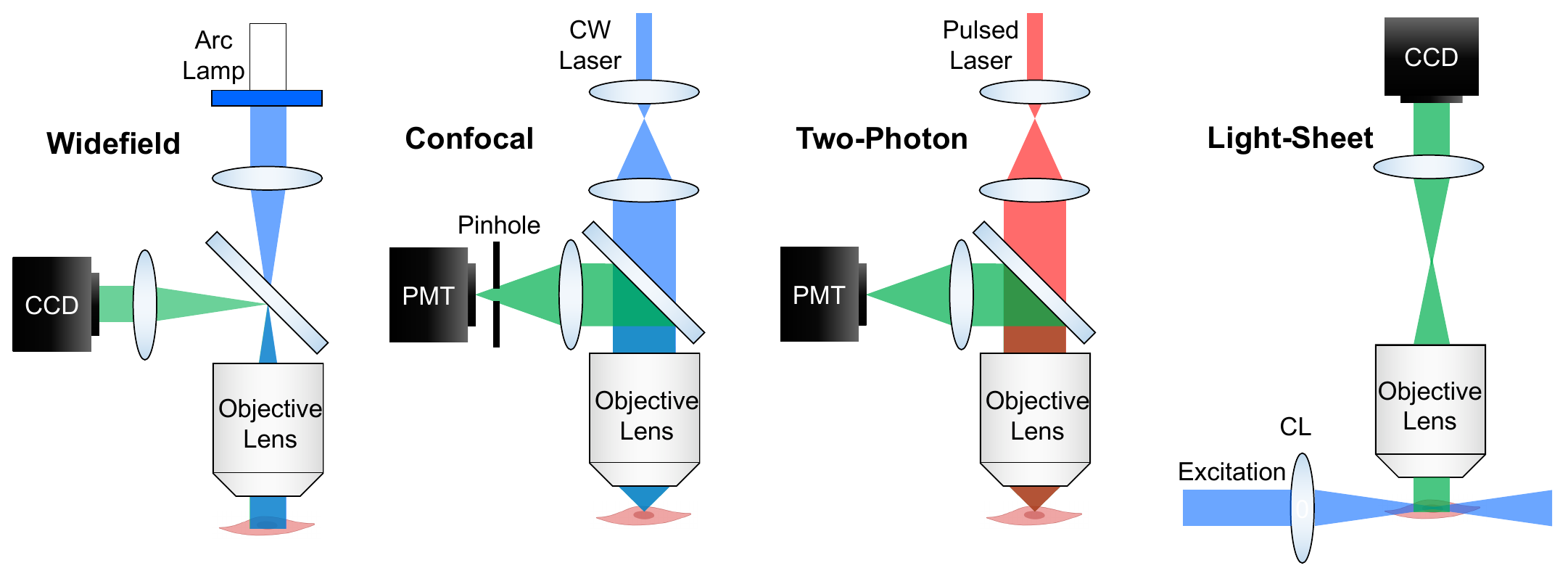}
	\caption{Schematics of wide-field, confocal, two-photon, and light-sheet microscopes (CCD, charge-coupled device; PMT, photomultiplier tube; CL, cylindrical lens). \cc{Blue, red, and green light paths indicate UV ($\lambda\approx 400$ nm) one-photon excitation, IR ($\lambda\approx 800$ nm) two-photon excitation, and visible ($\lambda\approx 500-650 $ nm) fluorescence emission, respectively. Wide-field microscopy images the sample on to a focal point array. Pinhole blocks the out-of-focus light in confocal microscopy. Point-scanning confocal and two-photon microscopy scan a focused excitation spot within the sample. In light-sheet microscopy, excitation and emission are in orthogonal directions, and the sample field is imaged to a focal plane array.}}
	\label{fig:modalities}
\end{figure}

Fluorescence microscopy is commonly implemented in wide-field \cite{wide-field}, confocal \cite{Confocal_Handbook, ConfocalMicroscopy}, multi-photon (two-photon) \cite{MPM, NonlinearMagic}, and light-sheet microscopy modalities \cite{lightsheet} (Fig.~\ref{fig:modalities}). Wide-field microscopy achieves high-speed imaging by simultaneously exciting fluorophores over a wide area and imaging the fluorescence emission onto an imaging array (e.g., CCD camera). However, the wide-field microscopy images are typically limited to surface or in vitro imaging as they cannot resolve depth and are susceptible to image blurring in scattering tissue. Confocal microscopy overcomes this limitation by using a pinhole that is confocal with the focal volume, thereby blocking the scattered light that would cause image blurring. \cc{However, such blocking of light significantly reduces the signal to noise ratio (SNR) (and, therefore, speed) and imaging depth, especially in highly scattering media.} Multiphoton microscopy (MPM), in turn, improves imaging depth in highly scattering media by exploiting the properties of nonlinear optics to only excite fluorophores within a diffraction-limited 3D volume in tissue and by collecting all emitted light without a pinhole. Since only ballistic photons contribute to the nonlinear optical excitation, MPM is resilient against the scattering of excitation light. Recently, light-sheet microscopy has become a useful tool for high-speed, depth-resolved, low-excitation level biological imaging by exciting a tissue cross-section and imaging the emission with an imaging array. This approach gives exceptional results in cell culture and transparent organisms but introduces technical limitations when employed in intravital imaging \cite{scape,andrew}. 

\subsection{Fluorescence lifetime imaging microscopy} \label{flim}
\begin{figure}[!t]
	\centering
	\includegraphics[width=0.8\linewidth]{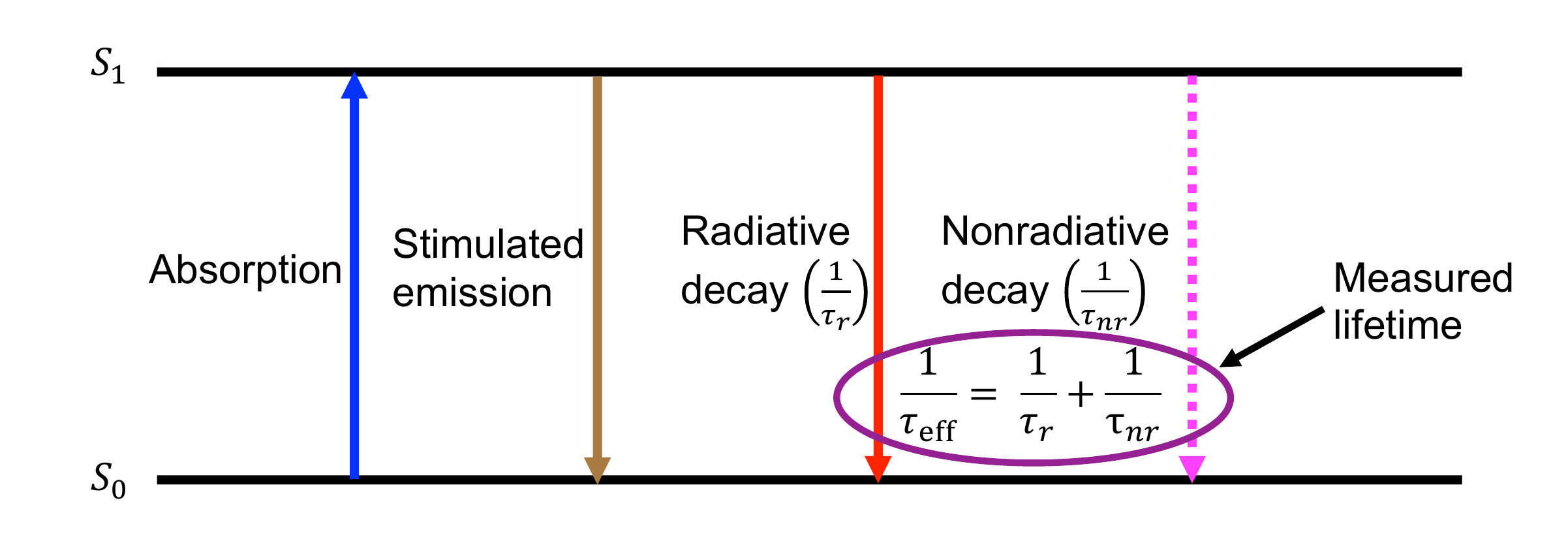}
	\caption{Jablonski diagram of fluorescence process. \cc{Absorption and stimulated emission rates are a function of photon flux and drive fluorophore between ground and excited states. Radiative decay and non-radiative decay process are spontaneous and characterized by their decay lifetimes, $\tau$. The radiative decay rate is considered a constant for a fluorophore and is represented by a solid line. The non-radiative decay rate is modulated by changes in the micro-environment and is represented with a dotted line. FLIM measurements commonly measure the effective decay rate, which is the sum of the radiative and non-radiative decay rates.}}
	\label{Jab_diag}
\end{figure}

The modalities in section \ref{fm} are fluorescence intensity-based techniques, which generate images by mapping fluorophore concentration (emission) to position in the sample. \cc{In addition to fluorescence intensity mapping, FLIM can provide additional information within the tissue compared to conventional intensity-based methods.} To understand what FLIM is measuring, we look to the fluorophore's rate equations. Fig.~\ref{Jab_diag} shows the Jablonski diagram illustrating the fluorescence generation process, where $S_0$ and $S_1$ are the ground and the first excited electronic states, respectively \cite{FLIM2007}. With the absorption of the excitation light by the fluorescence molecule, the electrons are excited from $S_0$ to $S_1$. \cc{The excited molecule returns to $S_0$ through a radiative (e.g., emitting fluorescence photons) or non-radiative (e.g., thermal relaxation) spontaneous process, or through a stimulated emission process. When the fluorophore emission is far below the fluorescence saturation emission rate, the spontaneous processes dominate and the stimulated emission contributions can be ignored.} The radiative decay rate is generally constant, while the local micro-environment can modulate the non-radiative decay rate. The effective measured lifetime $\tau_{\mathrm{eff}}$ is therefore dependent on both the radiative lifetime $\tau_{r}$ and the non-radiative lifetime $\tau_{nr}$. The fluorescence lifetime indicates the average time a fluorophore stays excited before relaxing and is typically in the range of nanoseconds \cite{Confocal_Handbook, FluorescenceLifetimeMeasurements}.

FLIM can be used to measure biologically important parameters such as ion concentrations, pH value, refractive index \cite{won2011precision}, the dissolved gas concentrations \cite{khan2014easily, khan2017silica, finikova2008oxygen}, and about the micro-environment in the living tissue \cite{sakadvzic2010two, ProteinInteractions, howard2013frequency}. Compared to conventional intensity-based fluorescence microscopy, FLIM is independent of fluorophore concentration, excitation power, and photobleaching, which are extremely difficult to control during most experiments \cite{FLIM_Optimization}.
\cc{Both fluorescence lifetime measurements and conventional intensity imaging accuracy are limited by shot noise, and FLIM implementation method could further limit fluorescence lifetime measurement accuracy by reducing the \enquote{photon economy} (i.e., the factor that SNR is reduced from an ideal, shot noise limited system). However, FLIM photon economy can approach the ideal shot noise limited SNR present in conventional intensity imaging when using time domain or Dirac-pulse frequency domain implementations \cite{josa, philip2003theoretical}.}
Moreover, since the lifetime is an intrinsic property of fluorescence molecules, FLIM is capable of separating fluorophores with overlapping emission spectra, which cannot be distinguished by traditional intensity-based imaging methods \cc{provided that there is a measurable difference in lifetime between the two fluorophores}. Another application of FLIM is F\"{o}rster resonance energy transfer (FRET), which \cc{can measure the distance between donor and acceptor molecules with 0.1 nm accuracy when within 10 nm of each other and with acceptable orientation \cite{FRET_ref, won2010referencing, hellenkamp2018precision_FRET}. FRET is extremely useful for applications such as protein studies in biological systems \cite{ma2014application_F1}, dynamic molecular events within living cells \cite{pollok1999using_F2}, and conformational changes in proteins \cite{rajoria2014flim_F3}.}

\begin{figure}[!t]
\centering
\includegraphics[width=0.9\linewidth]{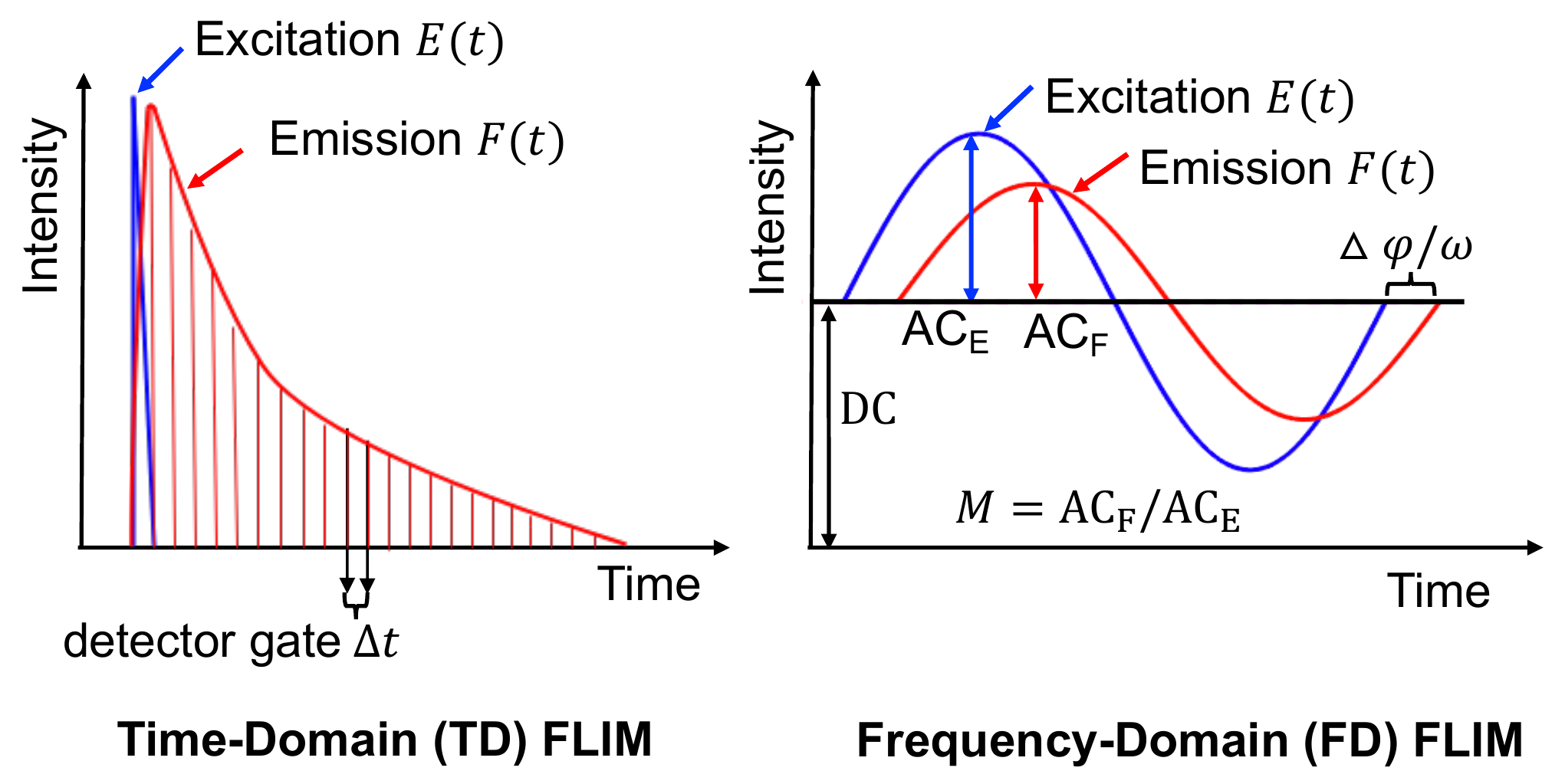}
\caption{Methods of fluorescence lifetime imaging microscopy. $E(t)$ and $F(t)$ are the excitation and emission signals, respectively. Left: time-domain FLIM implemented with time-gating; $\Delta t$ is the duration of the time gate. Right: frequency domain FLIM; $\mathrm{AC_E}$ and $\mathrm{AC_F}$ are the AC parts of the excitation and emission signals; $M$ and $\Delta \phi/\omega$ are the modulation degree change and phase shift, respectively, of the emission with respect to the excitation.} \label{td_fd_illu}
\end{figure}

FLIM data is acquired through one of two general methods: time-domain (TD) FLIM, such as time-correlated single-photon counting (TCSPC) \cite{TCSPC} and time-gating (TG)\cite{TimeGatingFLIM}; and frequency-domain (FD) FLIM \cite{SuperSensitivity,ProteinInteractions}, as shown in Fig.~\ref{td_fd_illu}. TD-FLIM methods measure the time between sample excitation by a pulsed laser and the arrival of the emitted photon at the detector. FD-FLIM methods, on the other hand, rely on the relative change in phase and modulation between the fluorescence emission and the periodically modulated excitation to generate the lifetime images.

In conventional TCSPC systems, the fluorescence lifetime is extracted from the measured fluorescence intensity decay by fitting the experimental TCSPC signal as either a mono-exponential function or multi-exponential function (for a mixture of fluorophores with different lifetimes). In a system consisting of fluorophores with different lifetimes, an effective lifetime ($\tau$) can be determined as the weighted average of the independent lifetimes. The TCSPC signal ($I(t)$) is the convolution between the system's instrument response function (IRF) and the weighted sum of fluorescent decay paths, as shown in Eq.~(\ref{TCSPCsignal}), where $\tau_n$ and $a_n$ are the lifetime and the corresponding weight of the $n\mathrm{-th}$ fluorophore population, respectively. 

\begin{equation}
\centering
    I(t) = \mathrm{IRF(t)} \ast \sum_n { a_n e^{-t/\tau_n}}.\label{TCSPCsignal}
\end{equation}

\begin{figure}[!t]
\centering
\includegraphics[width=12cm]{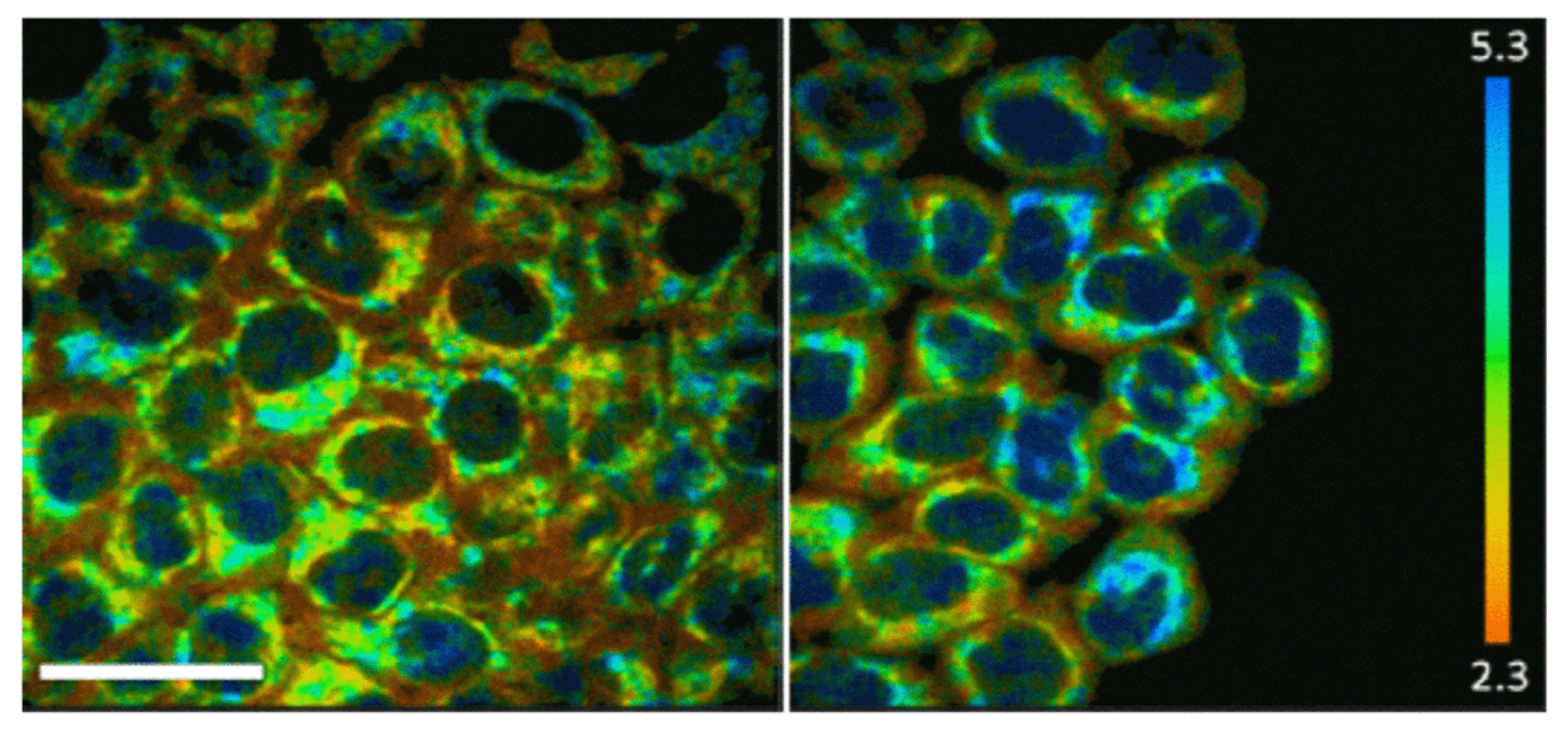}
\caption{NADH free/bound ratio mapping at different segments of a HeLa cell colony in a culture (taken on the 5th day of cell culture growth). The color coding reveals lower values of the $a_{1}/a_{2}$ ratio for the cells at the center (left) and higher values for the cells at the edge (right) of a colony, possibly indicating higher and lower metabolic activities correspondingly. Scale bar: 100 $\mathrm{\mu}$m. Data acquisition time: 700 s. [Adapted with permission from \cite{nadh}. Copyright (2009) American Chemical Society.]}\label{fig1_nadh}
\end{figure}

Extracting lifetimes and their weights from multiple decay paths allows one to find the ratio between two molecular populations with similar fluorescent emission spectra. For example, the functional/conformational states of nicotinamide adenine dinucleotide reduced (NADH) can be used to monitor oxidative phosphorylation, which is related to mitochondria oxidative stress \cite{nadh}. Fig.~\ref{fig1_nadh} shows how fluorescence lifetime changes with different cellular metabolic states in the NADH; hence by measuring fluorescence lifetime, one can extract the changes in the cellular metabolic state and thus accurately characterize normal and pathological states. Fig.~\ref{fig1_nadh} shows the extracted lifetime signal mapping by a bi-exponential model in the commercial software SPCImage v2.8 \cite{spcimage}. Here, cells in the left section indicate low $a_{1}/a_{2}$ ratio values, and cells in the right section indicate higher $a_{1}/a_{2}$ ratio values, which indicate higher and lower metabolic activities, respectively.    

However, the extraction of fluorescence lifetime with curve-fitting using Eq.~(\ref{TCSPCsignal}) in TD-FLIM data is computationally slow, and long acquisition time is needed to build up statistical confidence \cite{error_prop}. On the other hand, in time-gating FLIM systems (Fig.~\ref{td_fd_illu}), the emission fluorescence intensity is integrated with two or more time gates to extract the lifetime information of fluorophores with single or multiple exponential decays. However, an accurate lifetime estimation requires a large number of time gates (8 or more), which takes more time.

The above issues in TD-FLIM can be fixed using FD-FLIM techniques. In FD-FLIM, the excitation light is modulated at a frequency ($\omega$), and the emission fluorescence has the same frequency but a decreased amplitude and a shifted phase. Here, the amplitude modulation degree change is $m$, and the phase shift is $\phi$ \cite{howard2013frequency}. The lifetime in FD-FLIM can be extracted from either the amplitude change or the phase shift, which are independent of fluorescence intensity and fluorophore concentration. In FD-FLIM, for fluorophores with a single exponential decay, the lifetimes (at the $i\mathrm{-th}$ pixel) calculated from the modulation degree change ($\tau_m = \sqrt{1/[m_{i}^2-1]}$) and the phase shift ($\tau_\phi = \tan(\phi_{i})/w$) are identical \cite{josa}. For fluorophores of multiple exponential decays, the lifetimes calculated from the modulation degree change or phase shift are weighted averages of individual lifetime values. 

\begin{figure}[!t]
\centering
\includegraphics[width=8cm]{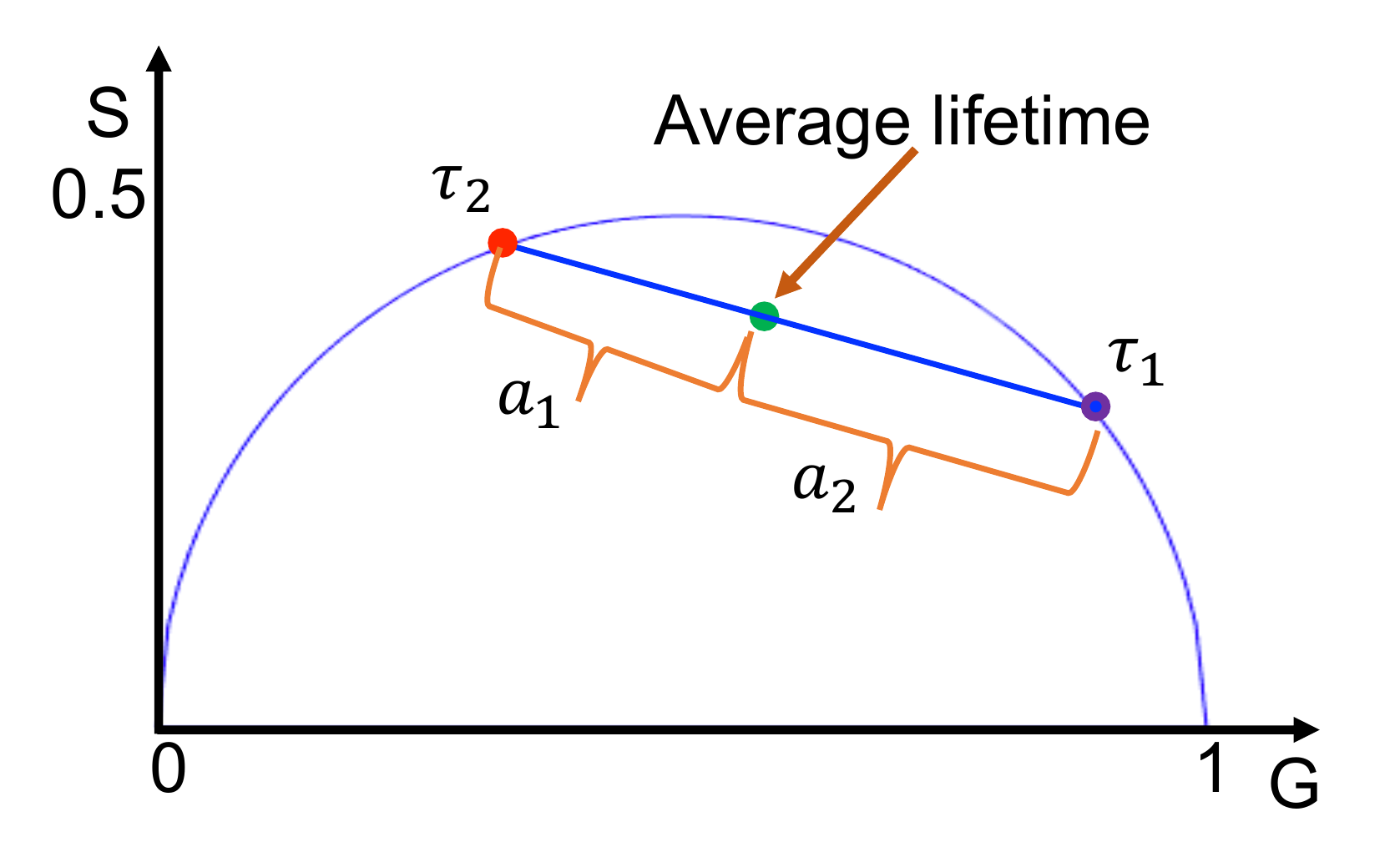}
\caption{Phasor plot for fluorophores with two exponential decays.}
\label{td_fd_expla}
\end{figure}

To simplify the interpretation and visualization of lifetime information, phasor plots can be used with raw TD- and FD-FLIM data \cite{redford2005polar}. As an example, Fig.~\ref{td_fd_expla} shows a simple phasor plot with two lifetime decays $\tau_1$ (short lifetime) and $\tau_2$ (long lifetime) and their weighted average. \cc{Phasor analysis is also useful analyzing molecules with multi-exponential decays, such as in commonly used FRET donor fluorophores \cite{ranjit2018fit, fereidouni2014phasor_c1, lou2019phasor_c2, chen2015method_c3, caiolfa2007monomer_c4}.}

Phasor plots are 2D representations of lifetime information, where the short lifetime is on the right side of the semi-circle, and the long lifetime is on the left. The coordinates along the x-axis ($G$) and y-axis ($S$) represent the real and imaginary values, respectively, of the phasor extracted from raw FLIM data. The two extremes of the semi-circle represent 0 and $\infty$ lifetimes with coordinates (1,0) and (0,0), respectively. For fluorophores with a single exponential decay, the phasor always lies on the semi-circle. For fluorophores with two exponential decays, the phasor lies on the line connecting the phasors corresponding to individual lifetimes. Similarly, for fluorophores with three exponential decays, the phasor is inside the triangle connecting all the three individual phasors.

The phasor plot is obtained using either TD- or FD-FLIM raw data \cite{flim_phasors}. For TCSPC data, the phasors can be acquired using the transformations $g_i = \int_0^\infty I(t)\cos(wt)dt/\int_0^\infty I(t)dt$ and $s_i = \int_0^\infty I(t)\sin(wt)dt/\int_0^\infty I(t)dt$, where $g_i$ and $s_i$ are the coordinates along the $G$ and $S$ axes, respectively, $w$ is the modulation frequency, and $I(t)$ is the TCSPC data at the $i\mathrm{-th}$ pixel. For FD-FLIM data, the phasors can be calculated using $g_i = m_i\cos(\phi_i), s_i =  m_i\sin(\phi_i)$, where $g_i$ and $s_i$ are the phasor coordinates, $m_i$ and $\phi_i$ are the modulation degree change and the phase shift of the emission with respect to the excitation at the $i\mathrm{-th}$ pixel. Regardless of whether the phasors are generated from TD-FLIM or FD-FLIM setups, the phasor coordinates are related to the lifetimes by $g_i = \sum_{k=1}^{n}a_{k,i}/[1+(w\tau_k)^2]$ and $s_i = \sum_{k=1}^{n}a_{k,i}w\tau_k/[1+(w\tau_k)^2]$, where $a_{k,i}$ is the intensity-weighted fractional contribution of the $k\mathrm{-th}$ fluorophore with lifetime $\tau_i$ at the $i\mathrm{-th}$ pixel and $\sum_{k=1}^{n}a_{k,i} = 1$ \cite{phasors}. The average lifetime at the $i\mathrm{-th}$ pixel is defined as the ratio of $s_i$ and $g_i$ with an additional factor of $1/w$.

Extraction of the fluorescence lifetime using the curve-fitting techniques is a complex process, and different software solutions are provided by different vendors\cite{spcimage, symphotime64, EasyTau2, flimfast, vistavision, flimfit, clip, li-flim, decayfit, cellSens}. All of these tools are computationally inefficient for real-time measurements. Many FLIM applications like FRET are quantitative and require data analysis in multiple dimensions, which can be overwhelming for existing software solutions mentioned above. To address these problems, we consider ML as one of the potential choices. ML is also helpful in interpreting the complex data and extracting the multi-dimensional data to a single exponent lifetime. The following sections will discuss the basic ideas and lifetime applications with ML.

\subsection{Machine learning} \label{ml}
Computing and interpreting TD-FLIM and FD-FLIM is computationally complex, slow, and leads to challenges in extracting biological interpretation from the raw data. ML approaches can be applied to accelerate imaging and aid interpretation significantly. Recently there has been a significant development in the processing of imaging techniques that mainly rely on ML methods. The ML methods have been extensively used in image processing and are model-free. A subset of ML is artificial neural networks (ANNs) that are inspired by biological neural networks and provide robust performance in applications such as systems identification and control \cite{control_ann}, classification, and medical image analysis \cite{ann_medical}. In medical image processing, ANN plays a vital role in the classification and segmentation of pre-cancer cells, resolution enhancement in histopathology \cite{histopath_ann}, super-resolution microscopy \cite{sr_ann}, fluorescence signal prediction from label-free images \cite{label_free_ann}, fluorescence microscopy image restoration \cite{restoration_ann}, and hyperspectral single-pixel lifetime imaging \cite{netflics1}.

In particular, deep neural networks (DNNs), a subset of ANN that uses a cascade of layers for feature extraction \cite{goodfellow2016deep}, mainly fully connected (FC) networks and convolutional neural networks (CNNs), have made significant progress in various imaging tasks such as classification, object detection, and image reconstruction. 

ML approaches have achieved unprecedented results; however, a few challenges are associated with them. First, the data-driven ML model requires a large training dataset, and such datasets are difficult to acquire. Second, training the ML model takes a significant amount of time (typically hours or days) and substantial resources (like high-performance graphical processing units).

\section{Overview of FLIM approaches using ML} \label{section2}
In this section, we consider publications that address fluorescence lifetime and its applications using machine learning. We divide the research articles into two main categories. First, those focusing on estimation of lifetime from input data (either in TD or FD) using ML. Second, those exploring the lifetime applications such as classification, segmentation using ML.

\begin{figure}[!t]
\centering
\includegraphics[width=14cm]{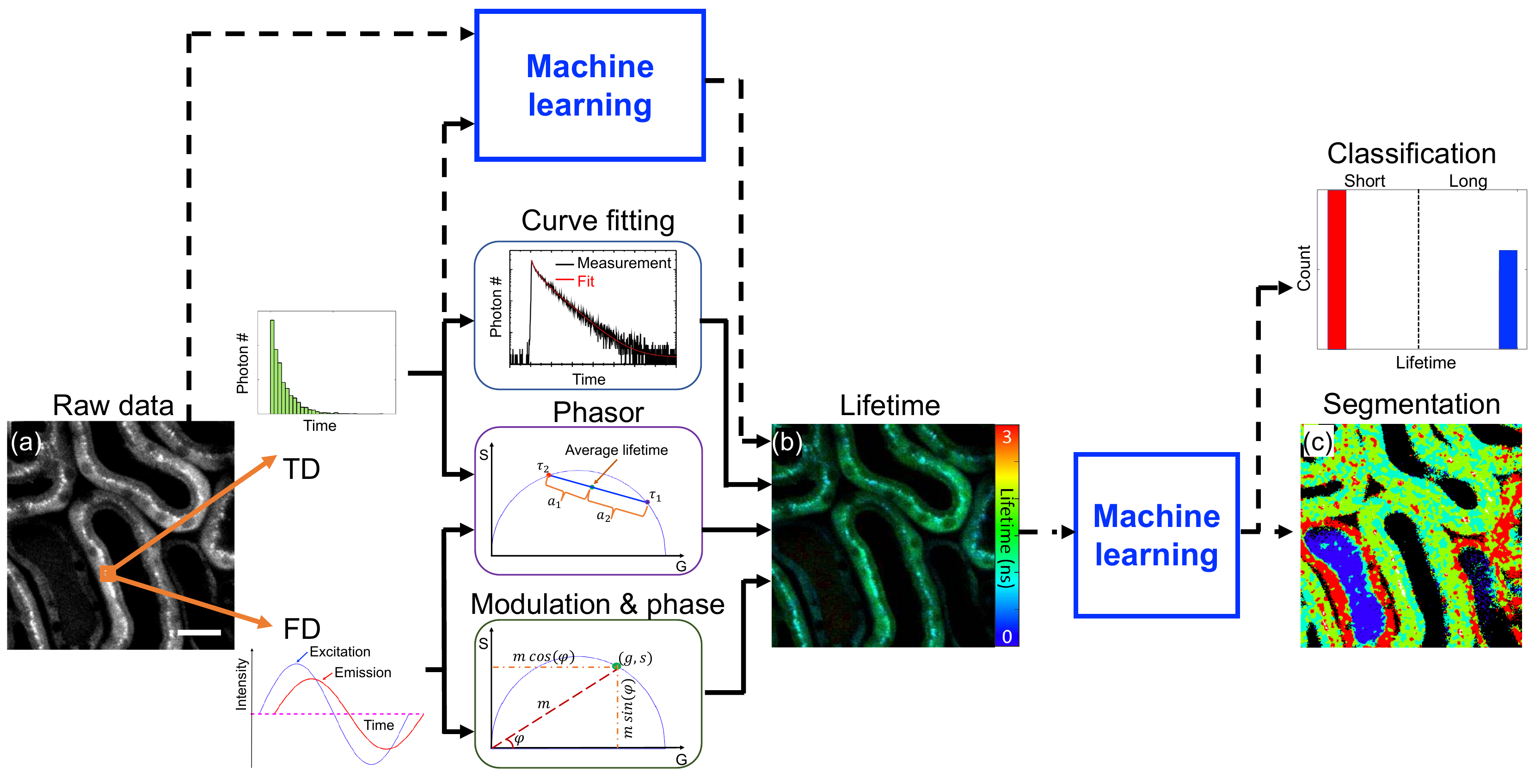}
\caption{Overview of FLIM approaches using ML: (a) raw data, (b) fluorescence lifetime image, and (c) lifetime image segmented with an ML approach. Solid line indicates conventional FLIM processing data flow; dashed line indicates ML FLIM processing data flow. \cc{In this flow chart, the ML block can include combinations of FC layers, CNNs, and unsupervised learning techniques for different applications, as described in the text. Microscope images are of intrinsic fluorophore imaging in an \textit{in~vivo} mouse kidney \cite{phasors}.} Scale bar: 20 $\mathrm{\mu}$m.}\label{fig_main4}
\end{figure}

Fig.~\ref{fig_main4} indicates the flow diagram of FLIM measurements using ML in a step-by-step process. \cc{First, the raw FLIM data is acquired using a TD- or FD-FLIM system.} Second, the lifetime can be extracted from either TD- or FD-FLIM measurements. In the time-domain, for each pixel, a histogram of the number of emitted photons is processed using curve fitting methods to estimate the lifetime parameters (lifetime values $\tau_n$ and their weights $a_n$, as in Eq.~(\ref{TCSPCsignal})). However, time-domain measurements (per-pixel) with curve fittings are computationally expensive and are not suitable for real-time measurements. Instead, one can efficiently solve for lifetime using ML methods. Similarly, ML methods can be used to estimate lifetime from TD measurements of the complete 3D stack ($x$, $y$, $t$). Another way to process this per-pixel histogram is by using phasors to get the average lifetime ($\tau_{\mathrm{avg}}$). In the frequency-domain, for each-pixel, changes in the modulation degree $m$ and phase $\phi$ are measured to calculate single exponent lifetime $\tau$. For multi-exponent lifetime values, $\tau_{\mathrm{avg}}$ can be estimated (for each-pixel) using the phasor approach. Lifetime estimation from input data using ML is discussed in section \ref{sec1}. 

Once the lifetime image is available, applications like classification and segmentation can be applied. Typical classification applications can be performed using basic statistics (mean and variance) on the lifetime. However, a threshold lifetime is required to classify lifetime images from these statistics. If training data is available, then a data-driven model can be employed to perform the same classification using ML methods. Classification and segmentation from lifetime using ML are discussed in sections \ref{sec2} and \ref{sec3}, respectively. Potential applications of lifetime information with ML and their proofs of concept are discussed in section \ref{potential_dir}.

\section{Estimation of FLIM images using ML}\label{sec1}
ML is well suited for estimating per-pixel fluorescence lifetime measurements from raw-data. Fluorescence lifetime can be estimated from TD-FLIM raw-data without computationally expensive curve fitting methods by using an artificial neural network (ANN) \cite{ann}.

\begin{figure}[!t]
\centering
\includegraphics[width=0.8\linewidth]{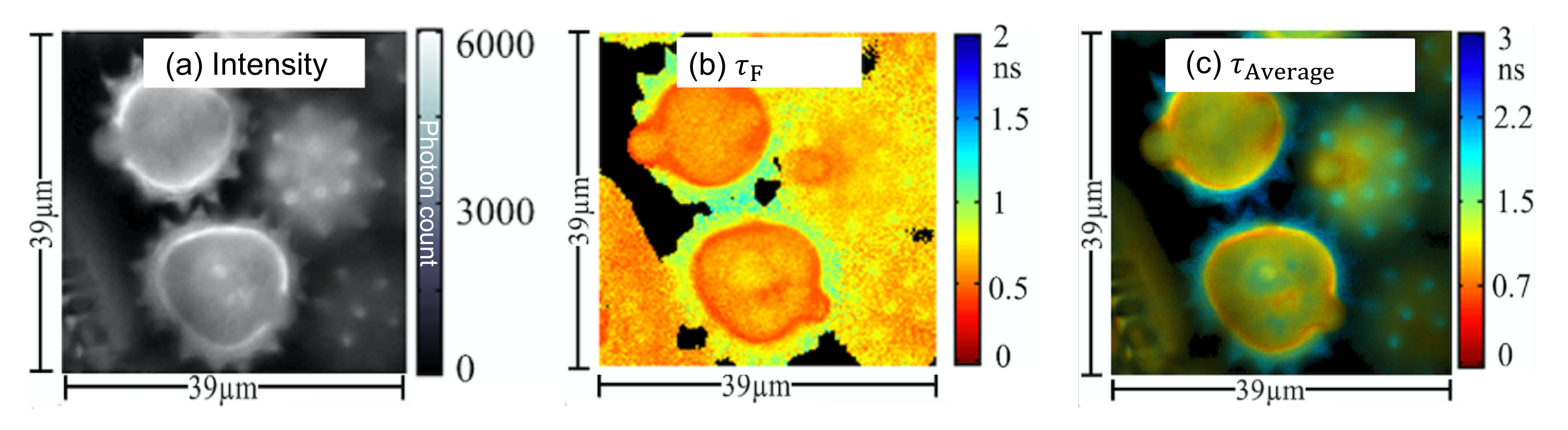}
\caption{(a) Intensity image, (b) short lifetime $\tau_F$ image generated by ANN and, (c) merged intensity and average lifetime $\tau_{\mathrm{Average}}$ image of daisy pollens. Adapted with permission from \cite{ann} $@$ The Optical Society.}\label{fig1_ann}
\end{figure}
 
Wu, et. al., used an ANN is used to approximate the function that maps TCSPC raw data as input onto the unknown lifetime parameters in \cite{ann}; Fig.~\ref{fig1_ann} displays this method applied to daisy pollens. Here an ANN model with two-hidden fully connected (FC) layers is trained with the TCSPC raw data from each pixel as input and its corresponding lifetimes (short and long lifetimes) and weights of the lifetime ratio as outputs. Fig.~\ref{fig1_ann}(a) shows the intensity of a daisy pollen in terms of photon count, and Fig.~\ref{fig1_ann}(b) shows the extracted short lifetime ($\tau_F$) using the trained ANN for the complete image. The estimation of the lifetime image (size of 256$\times$256) with the ANN method takes only 0.9 s, which is 180 times faster than the least-squares method (LSM) based curve-fitting technique, which takes 166 s. The weighted average of short and long lifetime results in an average lifetime image. The merged image of intensity and the average lifetime is shown in Fig.~\ref{fig1_ann}(c), where intensity and the fluorescence lifetimes are mapped to the pixels' brightness and hue, respectively. From \cite{ann} it was also observed that the LSM method failed to estimate the accurate lifetime parameters due to its sensitivity to initial conditions. The success rate for estimating accurate lifetime parameters with the ANN method was 99.93\%, and the LSM method was 95.93\%, which shows that the ANN method has improved estimation accuracy of lifetime from the TCSPC raw data compared to curve fitting tools \cite{ann}.

\begin{figure}[!t]
\centering
\includegraphics[width=12cm]{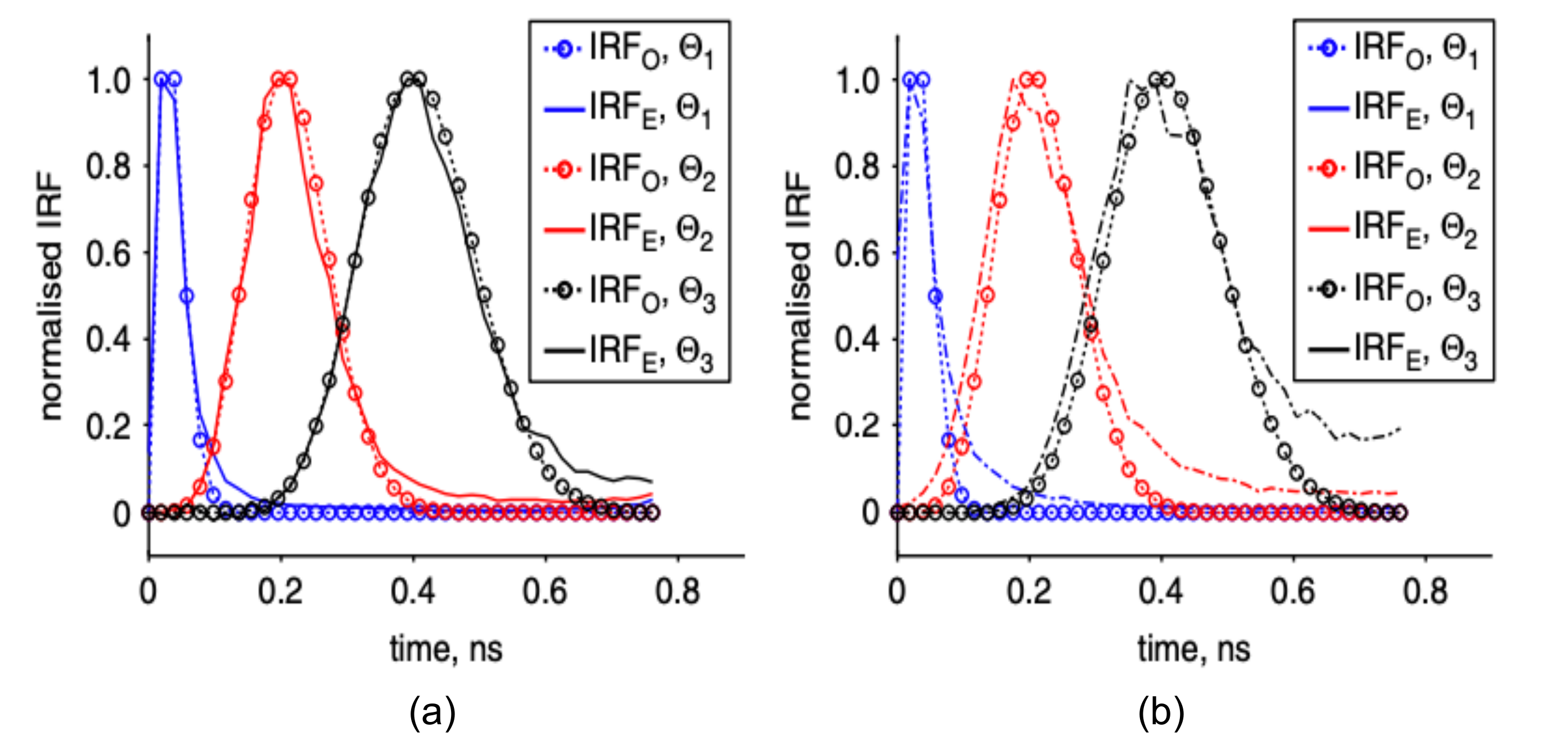}
\caption{Comparison of estimated IRF ($\mathrm{IRF_E}$) and original IRF ($\mathrm{IRF_O}$) with different lifetime (a) $\tau$ = 0.5 ns and (b) $\tau$ = 1.5 ns. The full widths at half maximum of IRFs, $\Theta$ where $\Theta_1$, $\Theta_2$, and $\Theta_3$ are 50 ps, 150 ps, and 200 ps respectively [Reuse permission is taken from the author \cite{em}].}\label{fig1_em}
\end{figure}

Another area where ML can improve data analysis/computing efficiency is estimating lifetime from a TCSPC system, including an IRF. \cc{For a TCSPC system, IRF must be measured throughout a set of FLIM measurements to account for variations in pulse shape during the experiment.} The deconvolution of IRF with TCSPC data provides lifetime information. However, measuring IRF requires additional steps before performing FLIM measurements. Many lifetime analysis tools ignore the IRF and use tail-fitting for estimation of a lifetime. ML can estimate the instrument response function $\mathrm{IRF(t)}$ in addition to single exponent lifetime $\tau$ from the TCSPC histogram data at every pixel with the Expectation-Maximization (EM) algorithm \cite{em}. Gao, et. al., demonstrated the EM to estimate the model parameters $\tau$ and IRF in \cite{em}. The input to EM is the histogram of TCSPC data, and the output is lifetime and IRF. Fig.~\ref{fig1_em} illustrates the results for the estimation of lifetime and IRF for different configurations. Thus, Gao and Li \cite{em} show that the IRF profiles could be extracted accurately using the EM method.

As we have seen, extracting lifetime from TCSPC data for each pixel is computationally inefficient even when applying ML approaches. Therefore, implementing ML techniques for 3D data ($x$, $y$, $t$) is challenging. Recently, Smith, et. al., proposed a novel method for estimating lifetime parameters on the complete 3D TCSPC dataset using convolutional neural networks (CNNs) in \cite{netflics2}. In \cite{netflics2}, the 3D CNN is trained using a synthetic data generator (from MNIST: handwritten digits dataset \cite{mnist}) to reconstruct the lifetime images ($\tau_1$ and $\tau_2$). In this network, input data is the spatial and time-resolved fluorescence decays as a 3D data-cube ($x$, $y$, $t$).

\begin{figure}[!t]
\centering
\includegraphics[width=9cm]{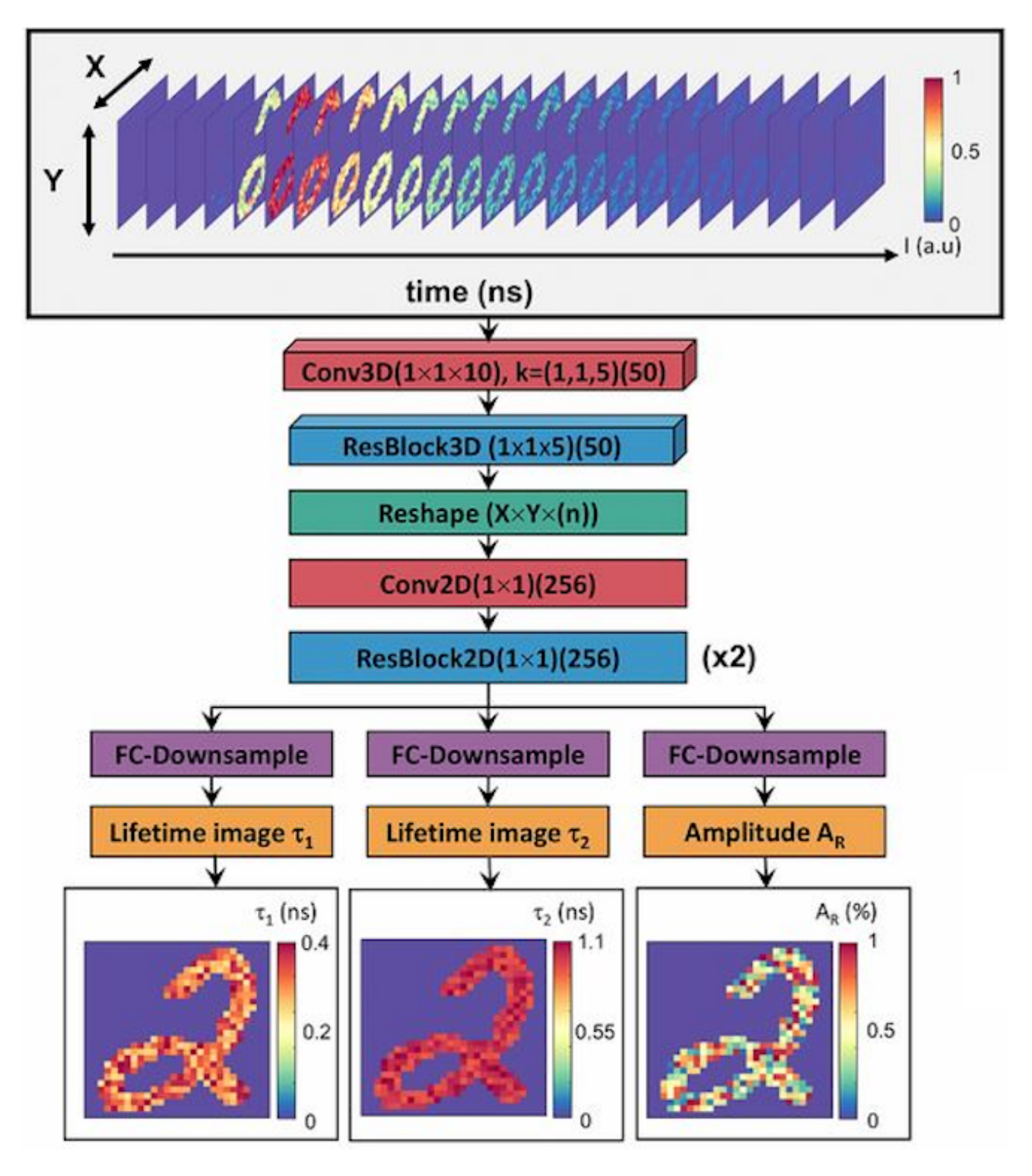}
\caption{Illustration of the 3D-CNN FLI-Net structure. During the training phase, the input to neural network was a set of simulated data voxels (which is the 3D data ($x$, $y$, $t$)) containing a TPSF at every nonzero value of a randomly chosen MNIST image. After a series of spatially independent 3D convolutions, the voxel underwent a reshape (from 4D to 3D) and is subsequently branched into three separate series of fully convolutional down-sampling for simultaneous spatially resolved quantification of 3 parameters, including short lifetime ($\tau_1$), long lifetime ($\tau_2$), and amplitude of the short lifetime ($A_R$) [Reuse permission is taken from the author \cite{netflics2}].} \label{fig1_netflics2}
\end{figure}

Fig.~\ref{fig1_netflics2} shows the architecture of fluorescence lifetime imaging-network (FLI-Net), which consists of a common branch for temporal feature extraction and subsequent split branches for the reconstruction of lifetimes (short $\tau_1$ and long $\tau_2$) and fractional amplitude of the short lifetime ($A_R$). In this architecture, the 3D convolutions (Conv3D) along the temporal dimension at each spatially located pixel are considered to maximize spatially independent feature extraction along with each temporal point spread function (TPSF). The use of a Conv3D layer with a kernel (which is a learning parameter during training) size of 1 $\times$ 1 $\times$ 10 reduces the potential introduction of unwanted artifacts from neighboring pixel information in the spatial dimensions ($x$ and $y$) during training and inference. Then a residual block (ResBlock) of reduced kernel length enables the further extraction of temporal information. After resolving the common features of the whole input, the network splits into three dedicated convolutional branches to estimate the individual lifetime-based parameters. In each of these branches, a sequence of convolutions is employed for down-sampling to the intended 2D image (short or long lifetime). Finally, Smith et al., \cite{netflics2} compared the results obtained from FLI-Net and the conventional least-squares fitting (LSF) method and observed that FLI-Net is accurate. 

Thus, we can solve a 3D CNN problem with TCSPC 3D data ($x$, $y$, $t$) as input and lifetime parameters as output, but memory is still a major challenge that must be addressed. During the training of this ML model with a TCSPC raw data, more memory (due to a large number of parameters) is required, which makes it difficult to process with a limited memory system. Performing image processing with massive data will be more challenging to perform in real-time; if the data is a 3D stack ($x$, $y$, $z$). For example, an image of size 256 $\times$ 256, with 256 time-resolved bins, each containing 10 bits (maximum value is 1023), has a size of 20.97 MB and if the data is a 3D stack ($x$, $y$, $z$), then the total size is in GB. To remedy this, Yao, et. al., combined compressive sensing (CS) with a deep learning framework to efficiently train the model in \cite{netflics1}.

\begin{figure}[!t]
\centering
\includegraphics[width=12cm]{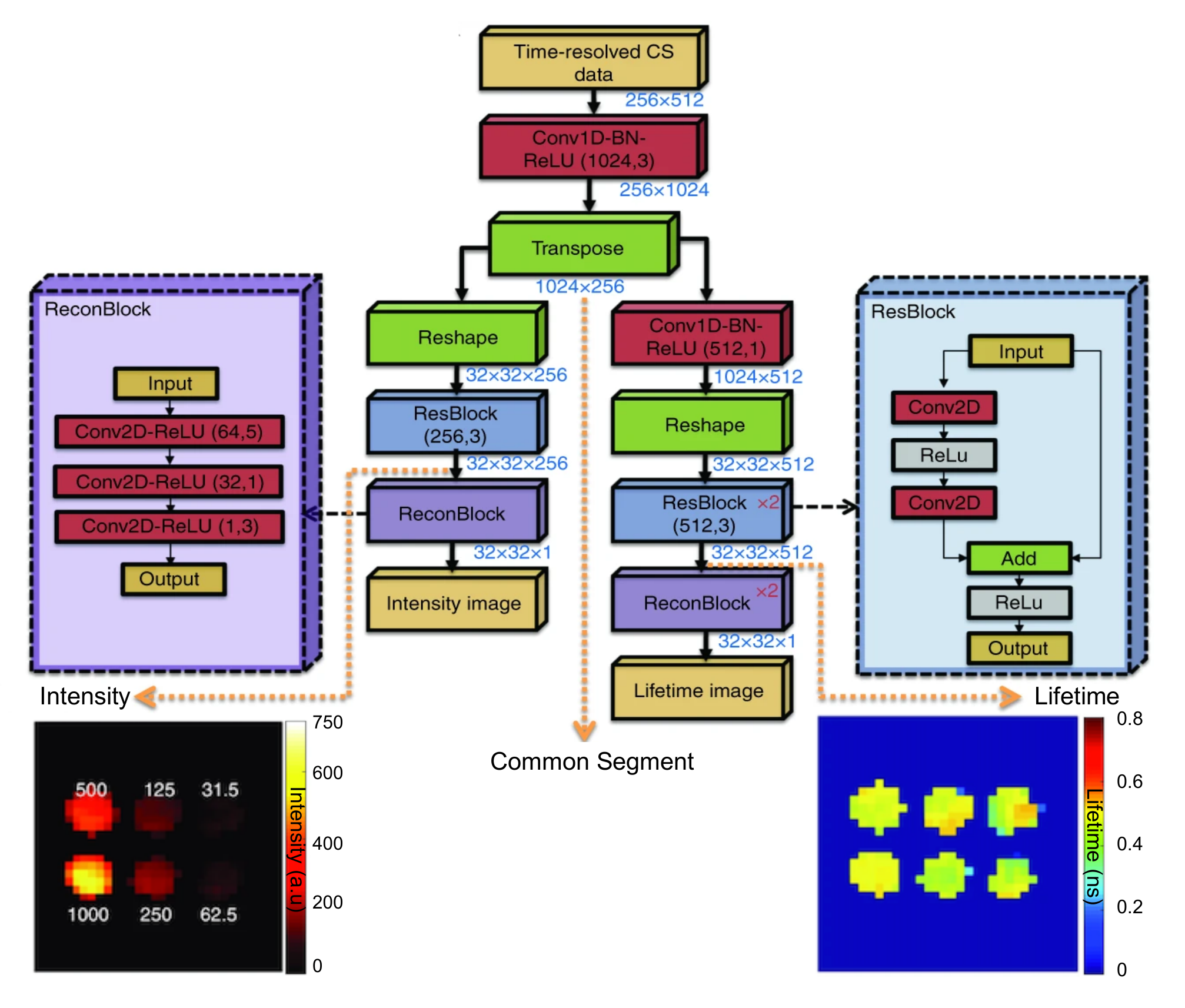}
\caption{The architecture of Net-FLICS and outputs at different endpoints of Net-FLICS (intensity and lifetime). Intensity and lifetime for the decreasing AF750 concentration from 1000 nM to 31.5 nM in six steps with true lifetime of 0.5 ns \cite{netflics1}.}\label{fig2_netflics1}
\end{figure}

In the CS approach, the input image data $I$ are related with compressive patterns (like Hadamard patterns $P$) such that $PI = S$, where $S$ is the CS data collected with patterns $P$. In this case, the intensity images are recovered using the CS-based inverse-solvers for each time gate, and the lifetime images are recovered using either curve-fitting tools or ML from the time-resolved intensity images. In multi-channel (different-wavelengths) CS data, the above procedure repeats for each channel. In \cite{netflics1}, the authors provide a ML architecture that extracts the intensity and lifetime images from the time-resolved CS data, as shown in Fig.~\ref{fig2_netflics1}. This architecture is defined as Net-FLICS (fluorescence lifetime imaging with compressive sensing data). Net-FLICS is a combination of two network designs: ReconNet \cite{reconnet} and Residual Network (ResNet) \cite{resnet}. The ReconNet is useful to reconstruct intensity images from random single-pixel measurements. ResNet is composed of one fully connected layer and six convolutional layers. Net-FLICS takes an array of size 256 $\times$ 512 as the input, which represents 512 CS measurements with 256 time gates, and outputs an intensity image and a lifetime image, both of which have sizes of 32 $\times$ 32, as the reconstruction prediction.  

In summary, Net-FLICS contains three main segments: (1) a shared segment that recovers the sparsity information from CS measurements; the first segment output represents the temporal point spread functions (TPSF) of each pixel; (2) an intensity reconstruction segment that contains one ResBlock and one ReconBlock; (3) a lifetime reconstruction segment that contains a 1D convolutional layer, two ResBlocks, and two ReconBlocks. Fig.~\ref{fig2_netflics1} shows the estimated intensity and lifetime from time-resolved CS data of Alexa Fluor dye (AF750, ThermoFisher Scientific, A33085), which is pH-insensitive over a wide molar range, for stable signals in imaging with six different concentrations starting from 1000 nM to 31.25 nM. The mean lifetime of the dye is 0.5 ns. As the concentration reduces, the number of detected photons also decreases. In the next section, we will explore the lifetime applications, like classification with ML.

\section{FLIM applications using ML: Classification} \label{sec2}
In this section, we enumerate the lifetime applications, surrounding classification of cells using ML. One application of lifetime is classifying pre-implantation mouse embryos' metabolic states as healthy or unhealthy. This estimation of embryo quality is especially helpful for $in~vitro$ fertilization (IVF) \cite{label}. Health monitoring can be achieved via invasive and non-invasive methods, but non-invasive methods significantly reduce the time and human resources necessary for analysis compared to invasive methods. Thus, estimating the quality of embryos using non-invasive methods is preferable, and extraction of lifetime information is an important step in these methods. In this non-invasive method, lifetime information is extracted from embryos at different time steps to monitor the cell culture at every stage. This lifetime information, combined with a phasor plot, shows embryos' growth in a specific way. Ma, et. al., demonstrated the phasor plot at different time steps of embryos correlates with embryonic development, and this phasor trajectory is called a development trajectory in \cite{label}. Here, the lifetime trajectory reveals the metabolic states of embryos.

With the ML algorithm, the trajectory of embryo health can be identified at an early stage by using a phasor plot of early samples. The ML algorithm is called distance analysis (DA) \cite{distance_analysis}, and it outputs the embryo viability index. In the DA algorithm \cite{distance_analysis}, the phasor plot is divided into four equidistance quadrants, and in each quadrant, a few statistics are measured (see \cite{distance_analysis} for details). Finally, the measured statistics are used to identify healthy and unhealthy results. Based on the embryos viability index (EVI) from the DA algorithm, the embryo classifies as either healthy or unhealthy. 

\begin{figure}[!t]
\centering
\includegraphics[width=14cm]{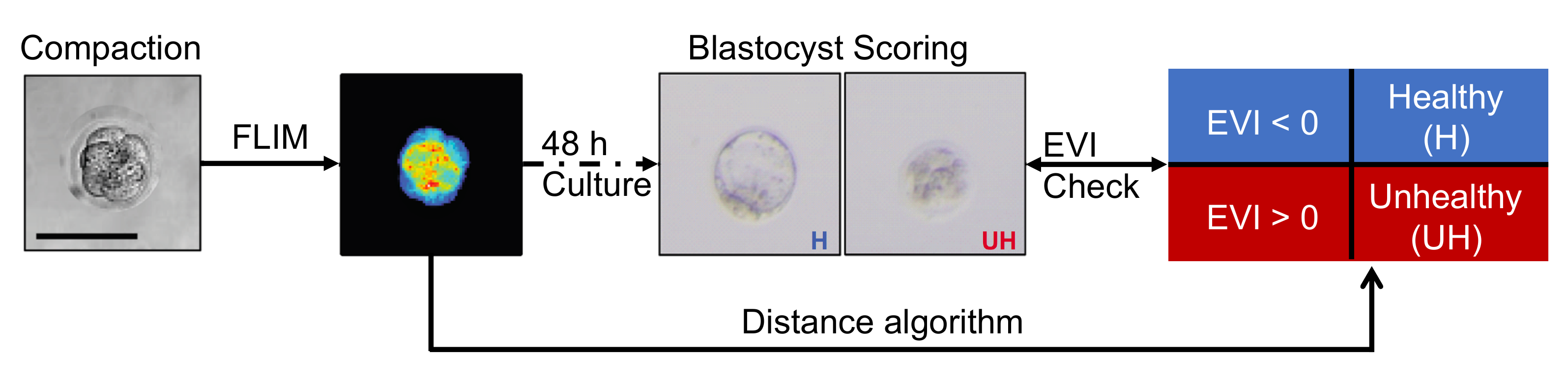}
\caption{Schematic of FLIM-Distance Analysis Pipeline to classify the mouse embryos. Scale bar: 100 $\mathrm{\mu}$m \cite{label}.}\label{fig1_labels1}
\end{figure}

Fig.~\ref{fig1_labels1} shows the flowchart for classification of the early-stage mouse embryos \cite{label}. First, embryos are collected at 1.5, 2.0, 3.0, 3.5, and 4.5 days post-fertilization and identified as 2-cell, Morula, Compaction, Early Blastocyst and, Blastocyst, respectively. Second, at the Compaction stage, the intensity image is passed through the FLIM setup to get the lifetime image. Here, lifetime information is generated using Compaction stage TCSPC data. Third, the generated phasor plot is passed through the DA algorithm to get the EVI. Finally, the Compaction stage embryo is classified based on its EVI. For the healthy (H) embryos, the EVI value is less than zero, and for the unhealthy (UH) embryos, the EVI value is greater than zero. Fig.~\ref{fig1_labels1} shows one healthy and one unhealthy embryo Blastocyst image identified by using the non-invasive lifetime and phasor approach with ML. 

\begin{figure}[!t]
\centering
\includegraphics[width=12cm]{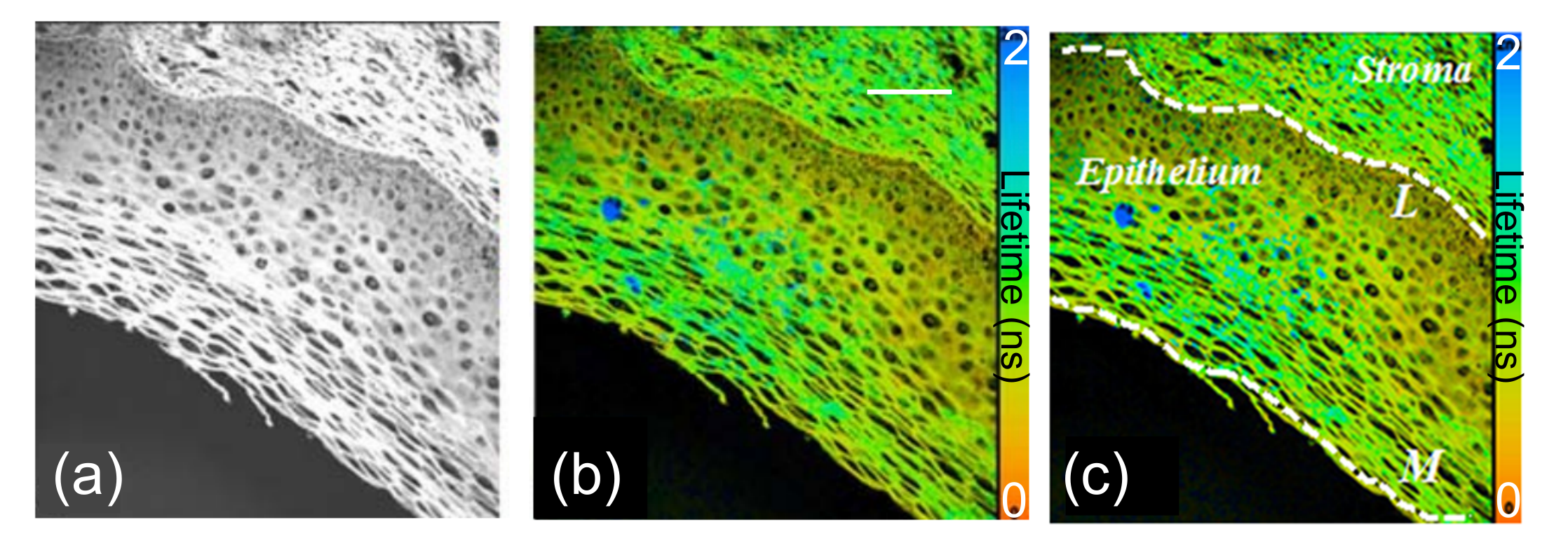}
\caption{Fluorescence (a) intensity and (b) lifetime images of a typical H \& E stained normal cervical tissue section. The region towards the right is the stroma, while the region on the left is the cell-rich epithelium separated by a visible boundary. (c) Lifetime images with markers where the basement membrane was marked out by white dashed line L and epithelium surface was delineated by white dashed line M. The color bars in (b), and (c) represent fluorescence lifetime on a scale of 0 (red) to 2 ns (blue). Scale bar: 100 $\mathrm{\mu}$m. [Reuse permission is taken from the author \cite{cancer1,cancer2}]}\label{fig1_cancer}
\end{figure}

ML can also be used to improve the quantitative diagnosis of precancer cells with lifetime information. For example, Carcinoma of the uterine cervix is the second-largest cause of cancer in women worldwide and is responsible for thousands of deaths annually \cite{cancer_intro}. Gu, et. al., imaged cervical tissue cells (a combination of the stroma and epithelium regions) using a fluorescence lifetime setup in \cite{cancer1}. Here the TCSPC data is captured, and lifetime information is extracted with a double exponent model at each pixel using SPCImage software. The extracted cells are classified as normal or cervical intraepithelial neoplasia (CIN) stage cells (from stage 1 to stage3) based on their lifetime information. Out of these normal cells, CIN1 stage cells are considered to be low-risk cells, while CIN2 and CIN3 stage cells are considered to be high-risk cells. A typical hematoxylin and eosin (H \& E) stained normal cervical tissue section is shown in Fig.~\ref{fig1_cancer}.

In these cells, stroma and epithelium regions are marked, as shown in Fig.~\ref{fig1_cancer} (c). In \cite{cancer1}, the longer lifetime ($\tau_2$) statistics (mean and standard deviation) of the epithelium region are calculated and considered as feature vectors of the ML model. In the model, lifetime feature vectors are passed through a single hidden layer neural network to classify the cells using extreme learning machine (ELM) algorithm \cite{elm}. ELM can accurately identify precancer cells in the epithelium region, and it can accurately classify precancer cells as low or high risk. Gu, et. al., separated the epithelium region into multiple layers, and each layer's feature vectors are measured in \cite{cancer2}. At each layer, feature vectors are used for classification with the ELM algorithm, to observe more accurate results. The mean lifetime of each layer is higher than the previous layer for normal cells, and for the precancer cells, the difference observed between each successive layer is less than in the healthy cells. In the next section, lifetime information applied to segmentation with ML is explained.

\section{FLIM applications using ML: Segmentation} \label{sec3}
In this section, we explore another application of FLIM, segmentation of cells using ML. In a given lifetime image containing two or more cells, it is essential to identify the locations of these cells and group them based on their lifetime values. However, measured lifetime at a pixel is an average value, and it cannot resolve the heterogeneity of fluorophores at that pixel since different fluorophore compositions (single fluorophore, multiple fluorophores) or excited state reactions could result in the same average lifetime measurements. Fortunately, different fluorophore compositions and possible excited state reactions can alter the phasor coordinates even if the average lifetime might be unchanged; therefore, the heterogeneity of fluorophores can be resolved with the phasor approach. 

In the phasor plot, pixels with similar fluorescence decays group together. This feature is useful for segmenting pixels based on the similarity of their fluorescence decays. Therefore, the lifetime segmentation technique is simplified using phasors by selecting adjacent phasor points and labeling them with different colors. However, this technique often leads to biased results.  

\begin{figure}[!t]
\centering
\includegraphics[width=12cm]{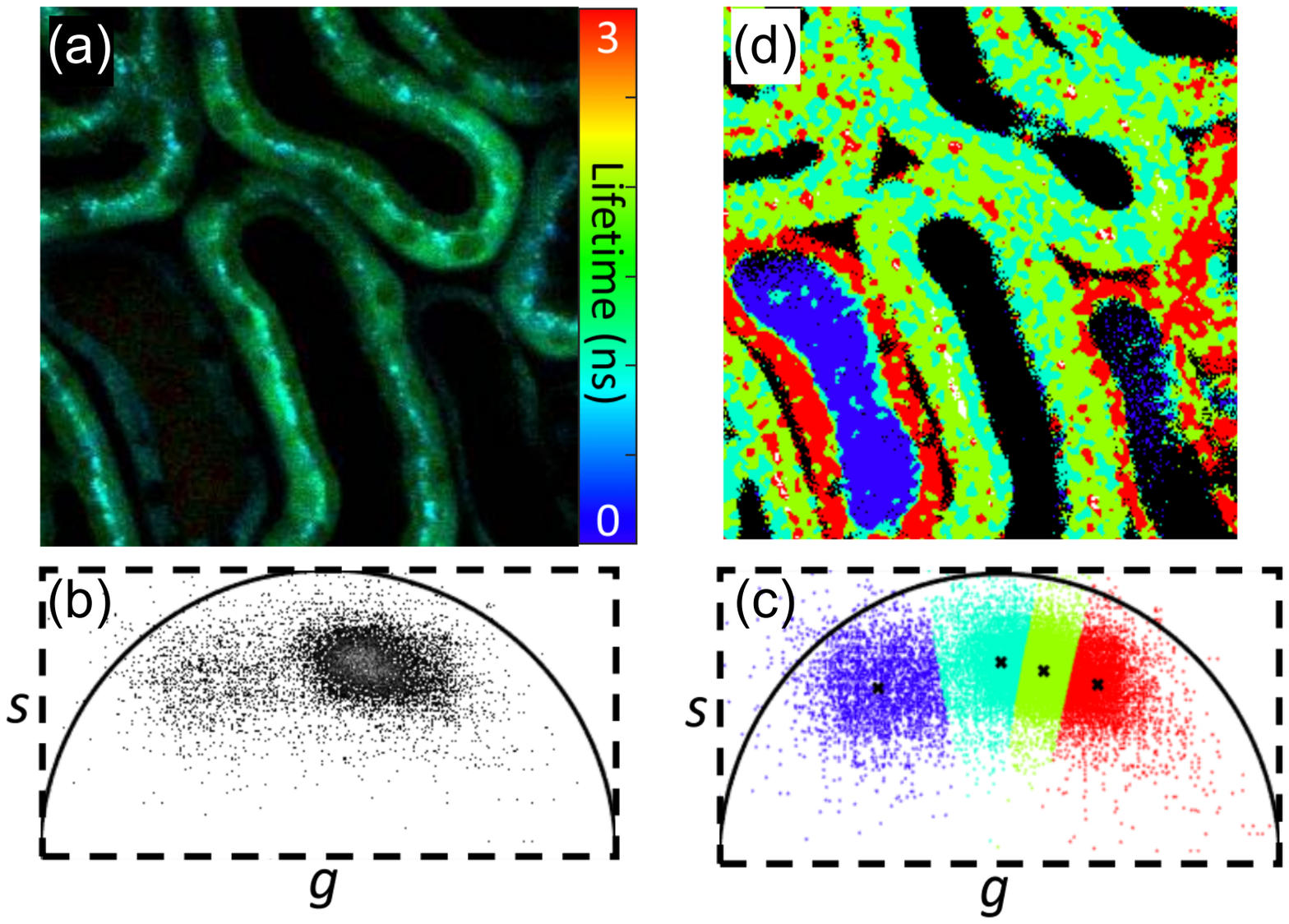}
\caption{Two-photon fluorescence lifetime image (a), phasor plot (b), K-means clustering on phasors (c), and segmented lifetime (d) of the kidney in a living mouse. Scale bar: 20 $\mathrm{\mu}$m. Adapted with permission from \cite{phasors} $@$The Optical Society.}\label{fig1_phasor}
\end{figure}

Zhang, et. al., demonstrated an unbiased method to segment the lifetime image using K-means clustering, an unsupervised ML algorithm in \cite{phasors}. Fig.~\ref{fig1_phasor} (a) shows the lifetime image of a living mouse kidney captured using our custom-built two-photon microscope \cite{instant_flim} and Fig.~\ref{fig1_phasor} (b) shows its phasor. By giving an estimated number, $K$, of different fluorophores present in the image, this approach separates its phasor plot into $K$ clusters using the K-means clustering algorithm. K-means clustering on the phasors shown in Fig.~\ref{fig1_phasor} (b) with four clusters is shown in Fig.~\ref{fig1_phasor} (c). The segmented lifetime image for each cluster with a different color in Fig.~\ref{fig1_phasor} (d) represents the different proximal tubules ($S_1$ and $S_2$) in the cells. K-means clustering technique is useful for segmenting phasors into groups without any bias. This segmentation method provides more consistent segmentation results and applies to both fixed and intravital 2D and 3D images. In the next section, the potential directions of FLIM using ML as well as their proofs of concept are discussed.

\section{Potential directions}\label{potential_dir}
\cc{The aforementioned ML-based FLIM applications (classification and segmentation) outperform conventional methods in terms of accuracy and computational time. Implementing ML techniques on fluorescence microscopy images has opened a considerable potential for research in the field of biomedical image processing. However, the potential uses of ML are not limited to the previously described techniques. To describe potential future directions, we demonstrate two proof-of-concept implementations of ML in FLIM for novel techniques. First, we demonstrate a framework for denoising fluorescence lifetime images using ML. Second, we illustrate estimating fluorescence lifetime images from conventional fluorescence intensity images.}

\subsection{Denoising of FLIM images using ML} \label{denoising_lifetime}
According to section \ref{section2}, it is clear that lifetime information is critical for classification and segmentation applications. However, the lifetime image ($\tau$) is noisy as it is obtained from the noisy raw data $I(t)$. Therefore, one of the future directions of the field is to perform fluorescence lifetime denoising using ML, similar to our previous fluorescence microscopy intensity denoising \cite{cvpr}. After the noise in the lifetime is reduced, the accuracy of classification and segmentation will increase. 

In \cite{phasors}, the lifetime is defined as the ratio of $S$ and $G$ (with an additional factor of $1/\omega$) at lock-in (modulation) frequency. Hence we can consider denoising $S$ and $G$ images separately, and the denoised lifetime becomes $\hat{S}/(\hat{G}w)$, where $\hat{S}$ and $\hat{G}$ are the denoised $S$ and $G$, respectively. We observe that denoising $S$ and $G$ images independently provide better results compared to direct denoising of lifetime images.

For image denoising, there are several methods proposed. One of the methods was discussed in our previous paper \cite{cvpr} to perform fluorescence intensity denoising using \enquote{Noise2Noise} convolutional neural network for the mixture of Poisson-Gaussian noise. In \cite{cvpr}, image denoising with ML outperforms all the existing denoising methods in two areas: accuracy (around 7.5 dB improvement in peak signal-to-noise ratio (PSNR)) and test time, which is less than 3 seconds using a central processing unit (CPU). The accuracy obtained in \cite{cvpr} is better since the ML model was trained with confocal, two-photon, and wide-field microscopic modalities which provide the most generalized model that can solve the denoising of Poisson noise, Gaussian noise, and a mixture of both.

\begin{figure}[!t]
\centering
\includegraphics[width=14cm,height=8cm]{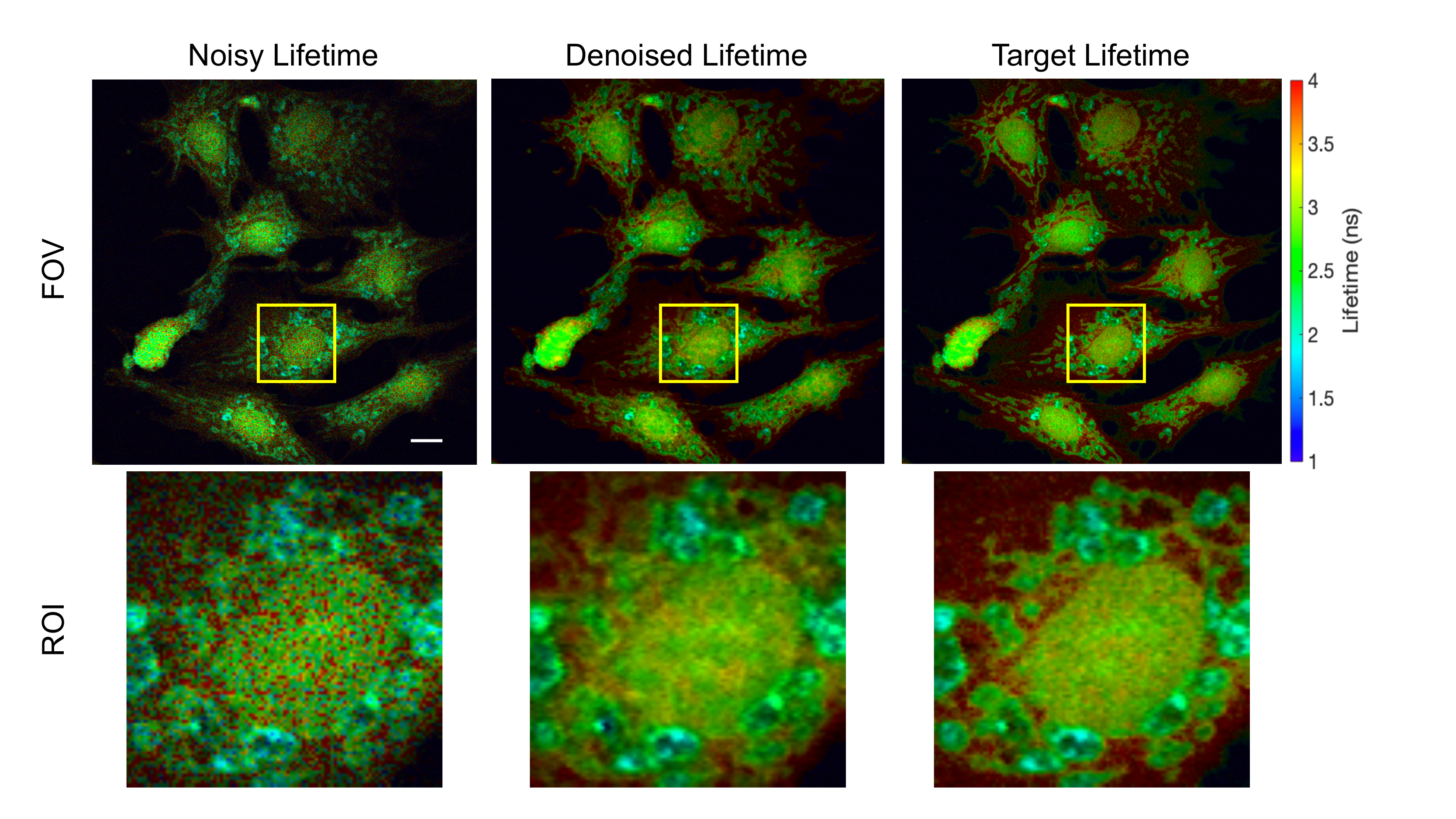}
\caption{Fluorescence lifetime denoising using ML. From left to right: noisy composite lifetime, denoised composite lifetime, target composite lifetime. Composite lifetime is the hue saturation value (HSV) representation of intensity and lifetime images together, where intensity and the fluorescence lifetimes are mapped to the pixels’ brightness and hue, respectively. The top row indicates a field of view (FOV) of 512 $\times$ 512 size, and the bottom row shows the region of interest (ROI) from the FOV (as shown in the yellow box) of size 100 $\times$ 100. The selected ROI indicates nuclei and mitochondria in green and light-blue colors, respectively. Scale bar: 20 $\mathrm{\mu}$m.}\label{pd_fig1}
\end{figure}

In this work, we implement the same Noise2Noise pre-trained model (trained for fluorescence intensity images) \cite{cvpr} to denoise the $S$ and $G$ images separately and obtain $\hat{S}$ and $\hat{G}$. In Fig.~\ref{pd_fig1} the denoised lifetime image is similar to the target image (where target lifetime is generated by averaging 50 noisy lifetime images within the same FOV) and much better than the noisy lifetime image. Raw lifetime images are captured with our custom-built two-photon microscope FLIM setup \cite{instant_flim} with an excitation wavelength of 800 nm, an image dimension of 512 $\times$ 512, the pixel dwell time of 12 $\mathrm{\mu}$s and pixel width of 500 nm. The samples we used are fixed BPAE cells [labeled with MitoTracker Red CMXRos (mitochondria), Alexa Fluor 488 phalloidin (F-actin), and DAPI (nuclei)] from Invitrogen FluoCells F36924. 

In Fig.~\ref{pd_fig1}, the noisy lifetime is generated from the noisy $S$ and $G$ images. These $S$ and $G$ images are passed through a pre-trained model in a plugin we developed for ImageJ \cite{Mannam:20} to get $\hat{S}$ and $\hat{G}$ images. The denoised lifetime is the ratio of $\hat{S}$ and $\hat{G}$. The improvement in PSNR in the denoised lifetime image is 8.26 dB compared to the noisy input lifetime image, which is equal to the average of 8 noisy images (in the same FOV)—thereby reducing the computation time by eight folds, instead of averaging more lifetime images for a better SNR. The denoised lifetime image (with high SNR) provides higher accuracy for classification and segmentation tasks. This work is a proof of concept for denoising lifetime images using a pre-trained model. If one is interested in obtaining better accuracy, the neural network has to be trained from scratch, which needs a large number of raw lifetime images as training data. As the acquisition of a large number of lifetime images is time-consuming, we show here the proof of concept using limited images and a pre-trained model.

\subsection{Estimation of FLIM images from intensity images using ML} \label{est_lifetime}
In section \ref{flim}, the estimation of lifetime using either TD or FD techniques is computationally expensive and requires additional hardware. One can construct a model with ML that reduces the computational time significantly. However, the ML model requires additional information such as the photon count histogram in TD or phase information in FD. The required hardware setup for acquiring this additional information is expensive for many research labs.
To address this issue, we propose another potential future direction to estimate fluorescence lifetime images from conventional fluorescence intensity images without either TD or FD setup. If one has a large training dataset (intensity and lifetime images), then a neural network model can be trained that maps the intensity to lifetime accurately.

\cc{As a proof of concept, in this section, we demonstrate the advantage of using machine learning for the estimation of fluorescence lifetime from a conventional intensity image using a limited training dataset and a specific biological model.}
In this work, we acquired a small dataset (intensity and lifetime image pairs) of a zebrafish that was two days post-fertilization and labeled with enhanced green fluorescence protein (EGFP labeled transgenic (Tg): sox10 megfp). The dataset size we use for training the network is 140 images, and the test data contains 16 images. Images in this $in~vivo$ dataset were captured with our custom-built two-photon microscope \cite{instant_flim} at different wavelengths and power levels to generalize the model. Thus, we train a NN model that maps the zebrafish intensity to zebrafish lifetime values.

\begin{figure}[!t]
\centering
\includegraphics[width=14cm]{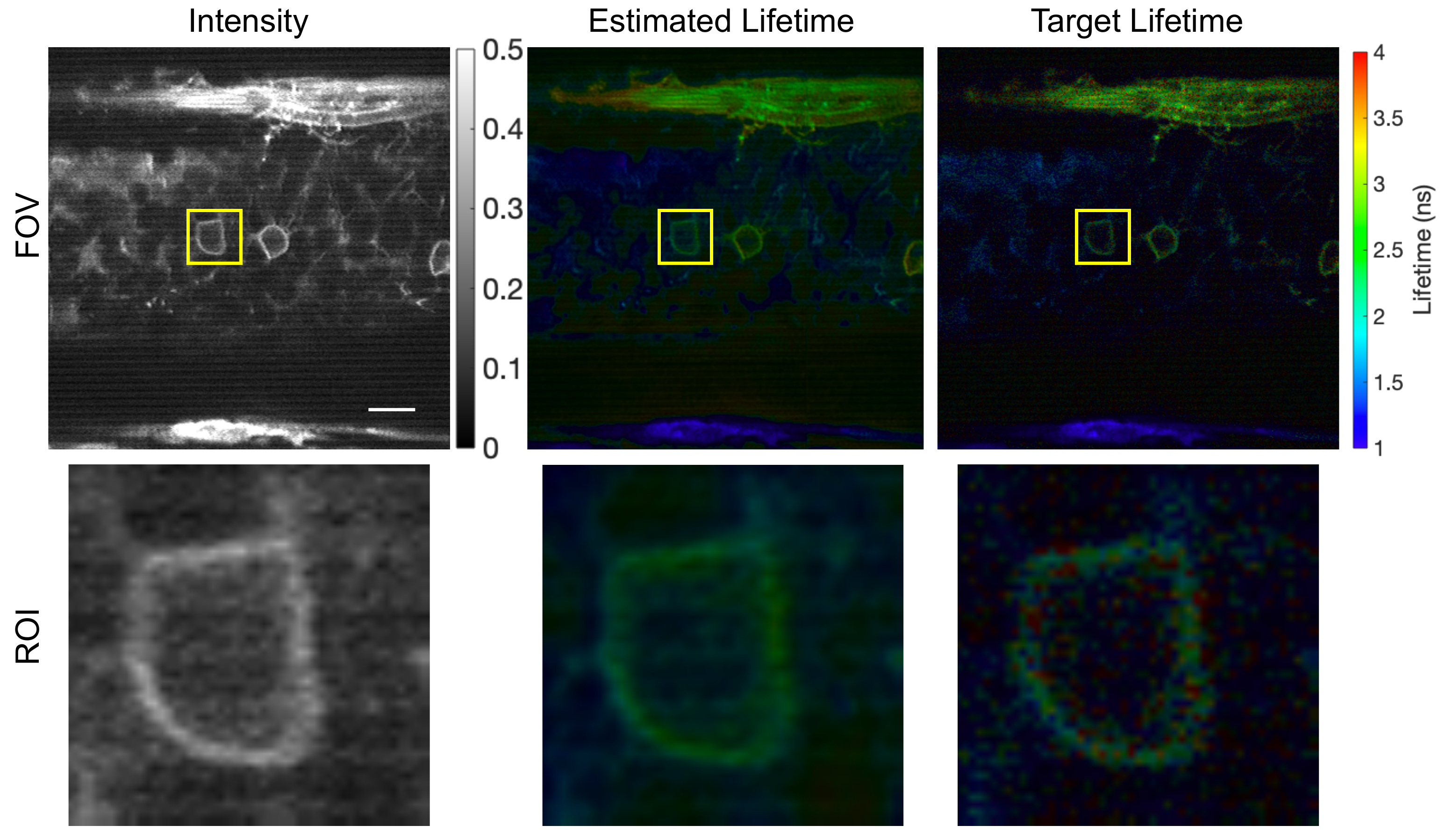}
\caption{Estimation of fluorescence lifetime from intensity using ML. From left to right: raw intensity image, estimated composite lifetime, target composite lifetime. Composite lifetime is the HSV representation of intensity and lifetime images together, where intensity and the fluorescence lifetimes are mapped to the pixels’ brightness and hue, respectively. The top row indicates the field of view (FOV) of size 360 $\times$ 360, and the bottom row indicates the region of interest (ROI) from the FOV (as shown in the yellow box) of size 50 $\times$ 50. Scale bar: 10 $\mathrm{\mu}$m.}\label{Fig2_PD}
\end{figure}

The estimated lifetime closely matches the target lifetime with a structural similarity index (SSIM) \cite{wang2004image_SSIM} of 0.9023, where \cc{SSIM value close to 1 indicates that the model prediction is good and also shows} how close the estimated and target images are matched. Here, we consider a famous neural network architecture (U-Net: which consists of an encoder and decoder structure \cite{mannam2020performance}) with an input of 256 $\times$ 256, encoded to the latent space of 8 $\times$ 8, and reconstructed back to a lifetime image of 256 $\times$ 256. We consider two training models with the target for one model to be lifetime and the target for the other model to be a composite lifetime. Here, composite lifetime is the HSV representation of intensity and lifetime images together, where intensity and the lifetimes are mapped to the pixels’ brightness and hue, respectively. We observe that the best lifetime estimation from intensity results occur when the composite lifetime, rather than the lifetime itself, is the target. Therefore, we choose the composite lifetime as the target to train our model. 

The model that we present has certain limitations. First, a large amount of data is required to train the model as lifetime information is unique to each molecule, and the trained model requires a massive collection of lifetime and intensity image pairs. Second, the model behaves differently for stained and unstained samples since intensity is mapped to a lifetime in the stained samples that are different, and therefore incorrect, in the unstained samples. Finally, retraining the NN takes more computational time and resources; therefore, in this work, we show a proof of concept with a limited image dataset and retrained NN model, as in Fig.~\ref{Fig2_PD}.

\section{Conclusions}
In this topical review, we have summarized the conventional lifetime extraction methods using both TD- and FD-FLIM techniques. Since the extraction of lifetime images is a complex process, ML stands out as a viable alternative to extract the lifetime at a faster rate with high accuracy. We have reviewed the novel applications of ML techniques to FLIM, such as a network for fluorescence lifetime imaging using compressive sensing data (Net-FLICS) and a 3D CNN fluorescence lifetime imaging network (FLI-Net) used for the lifetime extraction. Also, we have reviewed the applications of lifetime images such as classification and segmentation using novel ML techniques, including distance analysis (DA), extreme learning machine (ELM), and K-means clustering. Here, we have shown that the application of ML in FLIM significantly improves the computational speed and accuracy of the lifetime results.

Finally, we have proposed two additional potential directions for FLIM using ML approaches, including the denoising of fluorescence lifetime images and the estimation of lifetime images from conventional fluorescence intensity images. Both directions are presented with proofs of concept using datasets with a limited size. Our results demonstrate that ML can be one of the potential choices for future applications of lifetime images. For example, with a large dataset, a generalized model can be generated for the extraction of lifetime images from conventional fluorescence intensity images that significantly reduces the cost of a FLIM setup.

\section*{Disclosures}
\noindent The authors declare no conflicts of interest.

\section*{Funding Information}
This material is based upon work supported by the National Science Foundation (NSF) under Grant No. CBET-1554516.

\section*{Acknowledgments}
Yide Zhang’s research was supported by the Berry Family Foundation Graduate Fellowship of Advanced Diagnostics $\&$ Therapeutics (AD$\&$T), University of Notre Dame. The authors acknowledge the Notre Dame Integrated Imaging Facility (NDIIF) for the use of the Nikon A1R-MP confocal microscope and Nikon Eclipse 90i widefield microscope in NDIIF’s Optical Microscopy Core. The authors further acknowledge the Notre Dame Center for Research Computing (CRC) for providing the Nvidia GeForce GTX 1080-Ti GPU resources for training the neural networks using the Fluorescence Microscopy Denoising (FMD) dataset in TensorFlow.

\section*{References}
\bibliographystyle{unsrt}
\bibliography{references}
\end{document}